\documentclass[11pt]{article}

\usepackage{geometry}
\usepackage{currfile}
\usepackage{textcomp}
\usepackage{enumerate}

\usepackage{xcolor,soul}
\sethlcolor{yellow}

\usepackage[pdftex]{graphicx}
\DeclareGraphicsExtensions{.pdf,.jpeg,.png}
\graphicspath{{./.}}

\usepackage{fancyhdr}
\begin{document}

\title{\Large{Privacy-Preserving Claims Exchange Networks\\
for Virtual Asset Service Providers\\
~~\\
(Extended Abstract)}\\
~~}
\author{
\large{Thomas~Hardjono~~~Alexander~Lipton~~~Alex~Pentland}\\
\large{~~}\\
\large{MIT Connection Science \& Engineering}\\
\large{Massachusetts Institute of Technology}\\
\large{Cambridge, MA 02139, USA}\\
\large{~~}\\
\small{{\tt hardjono@mit.edu}~~{\tt alexlip@mit.edu}~~{\tt pentland@mit.edu}}\\
\large{~~}\\
}

\maketitle

\begin{abstract}
In order for VASPs to fulfill the regulatory requirements
from the FATF and the Travel Rule,
VASPs need truthful information regarding subjects,
such as originators, beneficiaries and other VASPs involved in a virtual asset transfer.
However, given that data about subjects are siloed in various organizations and institutions,
there needs to be a practical way for VASPs to obtain
information from these entities without direct access to the siloed data.
In this paper we describe the Open Algorithms approach
as a means for data holders to make insights about subjects available
to Claims Providers based on vetted algorithms.
A Claims Provider delivers signed claims to VASPs
regarding the relevant subject,
thereby relieving the VASP from having to deal with data, algorithms
and analytics.
We also propose a consortium arrangement for VASPs
to establish a Claims Exchange Network,
in which VASPs can deliver signed claims
(obtained from their Claims Providers)
and public-key information or certificates
to other VASPs in a secure and confidential manner.
~~\\
~~\\
{\bf Keywords}: virtual assets, blockchain technology, cryptocurrency, trust network, cryptography..

\end{abstract}

\newpage
\clearpage



~~\\

\section{Introduction}

Virtual asset service providers (VASP) face a data problem.
More specifically,
in order for VASPs to fulfill the regulatory requirements
from the FATF and the Travel Rule,
VASPs need access to truthful information regarding 
originators, beneficiaries and other VASPs involved in a virtual asset transfer.
However, getting access to data or information
-- regarding individuals and institutions involved in the asset transfer --
means that VASPs must also address the challenges
pertaining to data privacy and privacy-related regulations
such as the GDPR~\cite{GDPR} and CCPA~\cite{CCPA2018}.
On top of these issues,
in the past few years there has been
decreasing trust of consumers in institutions.
Negative reports regarding incidents of attacks on crypto-exchanges (e.g.~\cite{Dreyfuss2018})
compound this diminishing consumer trust.

We summarize these challenges as follows:
\begin{itemize}

\item	{\em The Travel Rule for virtual assets}:
The FATF Recommendation~15~\cite{FATF-Recommendation15-2018}
requires VASPs to retain information regarding the
originator and beneficiaries of virtual asset transfers.
This includes (i) originator's name;
(ii) originator's account number (e.g. at the Originating-VASP);
(iii) originator's geographical address, or national identity number, or customer identification number (or date and place of birth);
(iv) beneficiary's name;
(v) beneficiary account number (e.g. at the Beneficiary-VASP).

\item	{\em FinCEN compliance requirements}:
The FinCEN rules for anti-money laundering (AML) from 2014~\cite{FINCEN2014} requires that
customer due diligence (CDD) be performed for convertible virtual currencies~\cite{FINCEN2019}.

\item	{\em Decreasing trust of consumers in institutions}:
Over the last decade there has been a continuing decline in trust on the part of individuals
with regards to the handling and fair use of personal data~\cite{WEF2011,WEF2014}.
This situation has been compounded by the various recent reports of attacks and theft of data
(e.g. Anthem~\cite{Anthem2015}, Equifax~\cite{Equifax2017}).

\item	{\em Emergence of data privacy regulations}:
The enactment of the GDPR~\cite{GDPR} in Europe
has influenced the discourse regarding data privacy in other nations.
In the United States the state of California has enacted the
California Consumer Privacy Act (CCPA)~\cite{CCPA2018}.
Given the prominent role of data in the new digital economy,
the emergence of a US federal privacy act cannot be ruled out~\cite{Kerry2019a}.

\end{itemize}


\section{Virtual Assets and VASPs}
\label{sec:VirtualAssetsVASPs}

The {\em Financial Action Task Force} (FATF) is an inter-governmental body 
established in 1989 by the ministers of its member countries or jurisdictions.  
The objectives of the FATF are to set standards and promote effective 
implementation of legal, regulatory and operational measures 
for combating money laundering, terrorist financing and other 
related threats to the integrity of the international financial system.  
The FATF is a ``policy-making body'' which works to generate 
the necessary political will to bring about national legislative and regulatory reforms in these areas.

With the emergence of blockchain technologies, virtual assets and cryptocurrencies,
the FATF recognized the need to adequately mitigate the money laundering
(ML) and terrorist financing (TF) risks associated with virtual asset activities.
In its most recent Recommendation~15~\cite{FATF-Recommendation15-2018},
the FATF defines the following:
\begin{itemize}
\item	{\em Virtual Asset}: A virtual asset is 
a digital representation of value that can be 
digitally traded, or transferred, and can be used for payment or investment purposes. 
Virtual assets do not include digital representations of fiat currencies, 
securities and other financial assets that are already covered elsewhere in the FATF Recommendations.

\item	{\em Virtual Asset Service Providers} (VASP): Virtual asset service provider means 
any natural or legal person who is not covered elsewhere under the Recommendations, 
and as a business conducts one or more of the following activities or 
operations for or on behalf of another natural or legal person:
(i) exchange between virtual assets and fiat currencies; 
(ii) exchange between one or more forms of virtual assets;
(iii) transfer of virtual assets;
(iv) safekeeping and/or administration of virtual assets or instruments enabling control over virtual assets; and
(v) participation in and provision of financial services related to an issuer's offer and/or sale of a virtual asset.
\end{itemize}
In this context of virtual assets, transfer means to conduct a transaction 
on behalf of another natural or legal person that moves 
a virtual asset from one virtual asset address or account to another.
Furthermore, 
to manage and mitigate the risks emerging from virtual assets, the Recommendations states that
countries should ensure that VASPs are regulated for AML and Countering Financing of Terrorism (CFT) purposes, 
and licensed or registered and subject to effective systems 
for monitoring and ensuring compliance with the relevant measures called for in 
the FATF Recommendations.

\section{The Travel Rule and Customer Due Diligence}
\label{sec:travelrule}

One of the key aspects of the FATF Recommendation~15
is the need for VASPs to retain information regarding the
originator and beneficiaries of virtual asset transfers.
The implication of note~\cite{FATF-Guidance-2019} is that cryptocurrency exchanges and related VASPs
must be able to share the
originator and beneficiary information for virtual assets transactions.
This process -- also known as the {\em Travel Rule} --
originates from under the US Bank Secrecy Act (BSA - 31 USC 5311 - 5330),
which mandates that financial institutions deliver certain types of information
to the next financial institution when a funds transmittal event 
involves more than one financial institution.
This rule became effective in May 1996 and was issued by the 
Treasury Department's Financial Crimes Enforcement Network (FinCEN). 
This rule was issued by FinCEN concurrently with the new BSA record keeping rules
for funds transfers and transmittals of funds.

Given that today a virtual asset on blockchain
is controlled through the public-private keys bound to that asset,
we believe there are other information (in addition to the customer and account information)
that a VASP needs to retain in order to satisfy the travel rule~\cite{Hardjono2019b,HardjonoLipton2020a}:
\begin{itemize}

\item	{\em Key ownership information}: This is information pertaining to the 
legal ownership of cryptographic public-private keys.
When a customer (e.g. originator) presents their public key
to the VASP for the first time, 
there must be a ``chain of provenance'' evidence 
regarding the customer's public-private keys
which assures that the customer is the true owner.
Proof of possession of the private key (e.g. using a challenge-response protocol, such as CHAP (RFC1994)) 
does not prove legal ownership of the public-private key.

\item	{\em Key operator information}: 
This is information or evidence pertaining to the legal custody by a VASP
of a customer's public-private keys.
This information is relevant for a VASP which
adopts a key-custody business model in which
the VASP holds and operates the customer's public-private keys
to perform transaction on behalf of the customer.

\end{itemize}


	%
	%

In the 2014 FinCEN  Know Your Customer (KYC) requirements under the BSA~\cite{FINCEN2014},
the proposed rules  contained explicit customer due diligence (CDD)
requirements and included a new regulatory requirement to 
identify ``beneficial owners'' of customers who are legal entities.
It is worthwhile to note that the CDD requirements include 
{\em conducting ongoing monitoring to maintain and update customer
information} and to identify and report
suspicious transactions.
Collectively, these elements comprise the minimum
standard of CDD, which FinCEN
believes is fundamental to an effective AML program.

The FATF definition of virtual assets (``a digital representation of value that can be 
digitally traded, or transferred, and can be used for payment or investment purposes'')
means that VASPs -- like traditional financial institutions -- need to establish
an effective AML and CDD program in the sense of FinCEN~\cite{FINCEN2014,FINCEN2019}.
We believe that VASPs must additionally obtain and retain the originator/beneficiary
cryptographic key ownership information as a core part of monitoring 
the movement of virtual assets.

\section{Related Work: Identity Claims Model}
\label{sec:ClaimsModel}

The problem of customer identification, on-boarding and due diligence
is not unique to VASPs, and has been a challenge for 
Internet service providers generally (i.e. online merchants) since the late 1990s.
The promise of Internet-based services (versus traditional brick-and-mortar shops)
was that of an increase in transaction efficiency, lower costs and better convenience for the user.
However, as the past two decades of Internet services has shown,
the problem of consumer identification and authentication is not trivial and
is closely related to the problem of data and consent-based access~\cite{WEF2011,hardjono2019a} to
personal data pertaining to the consumer (data subject)~\cite{GDPR}.

\begin{figure}[!t]
\centering
\includegraphics[width=0.6\textwidth, trim={0.0cm 0.0cm 0.0cm 0.0cm}, clip]{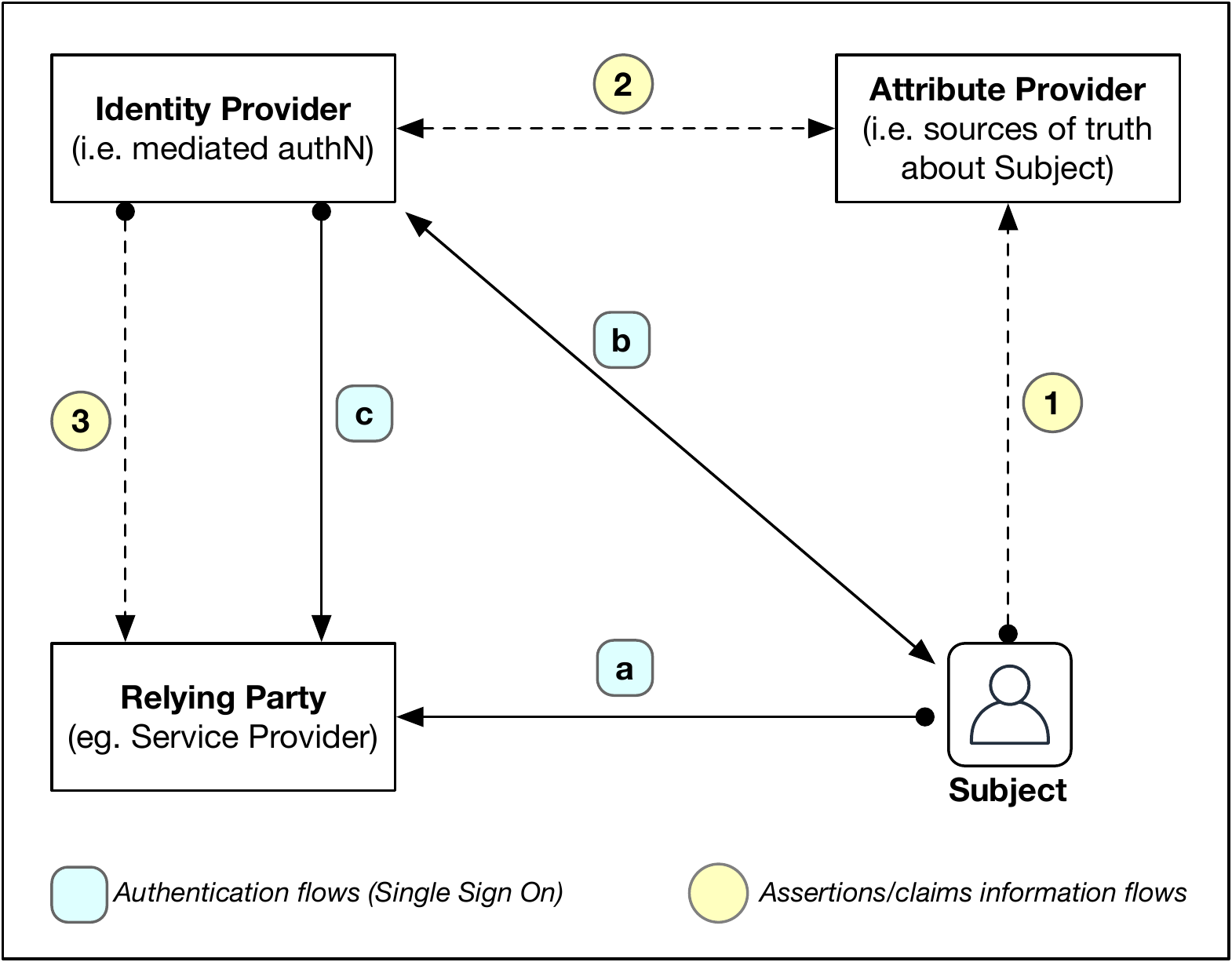}

	%
	%
\caption{The SAML Mediated Authentication \& Attributes Delivery flow}
\label{fig:idpbasic}
\end{figure}

\subsection{Attributes in the SAML2.0 Model} 

Online services today employ {\em Identity Providers} (IdP)
as means to provide mediated authentication of the user (subject)
on behalf of the online Service Providers (SP), such as online merchants.
The Service Providers are reliant on the authentication-event outcome of the IdP,
and as such they are referred to also as the Relying Party (RP).
This is referred to as {\em Web Single Sign-On} (Web-SSO)
for browser-based user interactions~\cite{SAMLcore}.
The typical consumer-facing IdP issues an identifier (e.g. email address) and 
manages the credentials of the user (e.g. change password). 
When the user seeks to access services offered by the Service Provider, 
the user is temporarily redirected by the SP to the IdP for authentication. 
If the authentication is successful, 
the IdP issues an authentication-token (e.g. SAML2.0 tokens, Kerberos tickets)
which can then be validated by the Service Provider.
The IdP and the Service Provider typically have a business relationship
that provides the foundation of trust between them.

Figure~\ref{fig:idpbasic} illustrates the basic mediated authentication flows through the IdP
in steps (a)--(c).
After the IdP provides the Service Provider (Relying Party)
with evidence of successful authentication in Step~(c),
the Service Provider now requires factual
information or {\em attributes} (assertions or claims) about the user (subject).
Here, one approach defined by the SAML2.0 specifications~\cite{SAMLcore}
is for the IdP to inquire to a special entity called the {\em Attribute Provider} (AtP)
to furnish the IdP with attribute assertions or claims about the subject.
In other literature (e.g.~\cite{Cameron2005identity}), the Attribute Provider
is also known as the {\em Claims Provider} (CP).
Thus, in Step~(1) of Figure~\ref{fig:idpbasic}
the subject provides consent or authorization for the Attribute Provider
to release information to the IdP in Step~(2).
The IdP forwards the assertions or claims in Step~(3) to the Service Provider.
Alternative flows are possible,
such as when the signed claims are delivered to the SP through the Subject.


From the VASP perspective,
the flows in Figure~\ref{fig:idpbasic} provide the rudimentary mechanism
for a VASP to obtain customer information in the form of signed claims.
Thus, the VASP is the relying party because
it is reliant on the Attribute Provider (Claims Provider)
to furnish it with information about the subject
seeking the services of the VASP (e.g. subject request transfer
of virtual assets).
Note, however,
that the authentication-token paradigm of~\cite{SAMLcore}
does not address {\em how} information about
a subject can be obtained or derived.

\subsection{Authorization to Access Protected Claims in {UMA2.0}  } 

The rise in mobile devices in the past decade
required a different authorization model than the Web-SSO of the 1990s.
In particular, many mobile applications (i.e. apps) require
on-going access to the user's online resources
(e.g. calendar, photos, email account, etc.)
that are distributed throughout the Internet
at different service providers.
Access to resources is needed
even when the user is not currently using the mobile app (e.g. background sync of email, calendar, etc.). 
Thus, a token-based authorization model emerged
based on the {OAuth2.0} framework~\cite{rfc6749}
which permitted the user to ``authorize''
mobile apps to continue to access the user's online resources,
and which provided an automatic ``refresh'' token that could be obtained
by the mobile app from the authorization server
without prompting the user.

The token-based authorization model of {OAuth2.0}~\cite{rfc6749}
was subsequently extended by the {\em User Managed Access} (UMA) architecture~\cite{UMACORE1.0,UMACORE2.0}
starting in early 2009.
The UMA effort recognized from the start that indeed consumer data
was dispersed throughout the Internet
and that if a consistent consent-based approach (such as that proposed by the GDPR~\cite{GDPR})
was to be implemented,
a {\em federated authorization model} was required~\cite{hardjono2019a}.
In this model,
personal data, claims and other information about the user (subject)
stored at various ``resource servers''
throughout the Internet could be accessed by a third-party
only if the requesting party first obtained consent from the user
through the UMA protocol.
Thus, the UMA approach specifically recognized the reality
that data about tens of millions of users are today
in the possession of various {\em data providers}
-- financial institutions, telecom operators, health organizations, 
social media platforms, email providers, etc. --
and that a protocol to implement decentralized consent management was needed.

\subsection{Consent for Execution of Algorithms to derive Claims}

An important aspect of the {OAuth2.0} framework~\cite{rfc6749} and of 
the UMA architecture~\cite{UMACORE1.0,UMACORE2.0}
is the absence of any mechanism to express {\em consent policies} (access rules) on the part of the user.
Both {OAuth2.0} and UMA saw consent expression languages 
as out-of-scope for the technical design of the resulting protocols.

Given the prevalent practice in industry of re-selling consumer
data to ``data brokers'' -- what Tim Cook from Apple refers to as the ``shadow market''~\cite{Cook2019} --
we believe that a safer approach is needed that disincentives the copying (export) of consumer data
from one institution to another.
We refer to this approach as {\em open algorithms} (OPAL)
based on a number of privacy-preserving principles (discussed further below in Section~\ref{sec:opal}).

Following from the open algorithms principles,
the work of~\cite{HardjonoPentland2018b-SHORT} has proposed
the notion of {\em consent for the execution of algorithms} from the user (subject).
Here, when a subject provides consent,
the default interpretation of ``consent'' is that of 
the data provider running a specific algorithm on the subject's data (in the repository, without exporting it).
The algorithm must be vetted by experts, published
and explained in lay-language to the subject (i.e. informed consent~\cite{GDPR}).

\begin{figure}[!t]
\centering
\includegraphics[width=0.6\textwidth, trim={0.0cm 0.0cm 0.0cm 0.0cm}, clip]{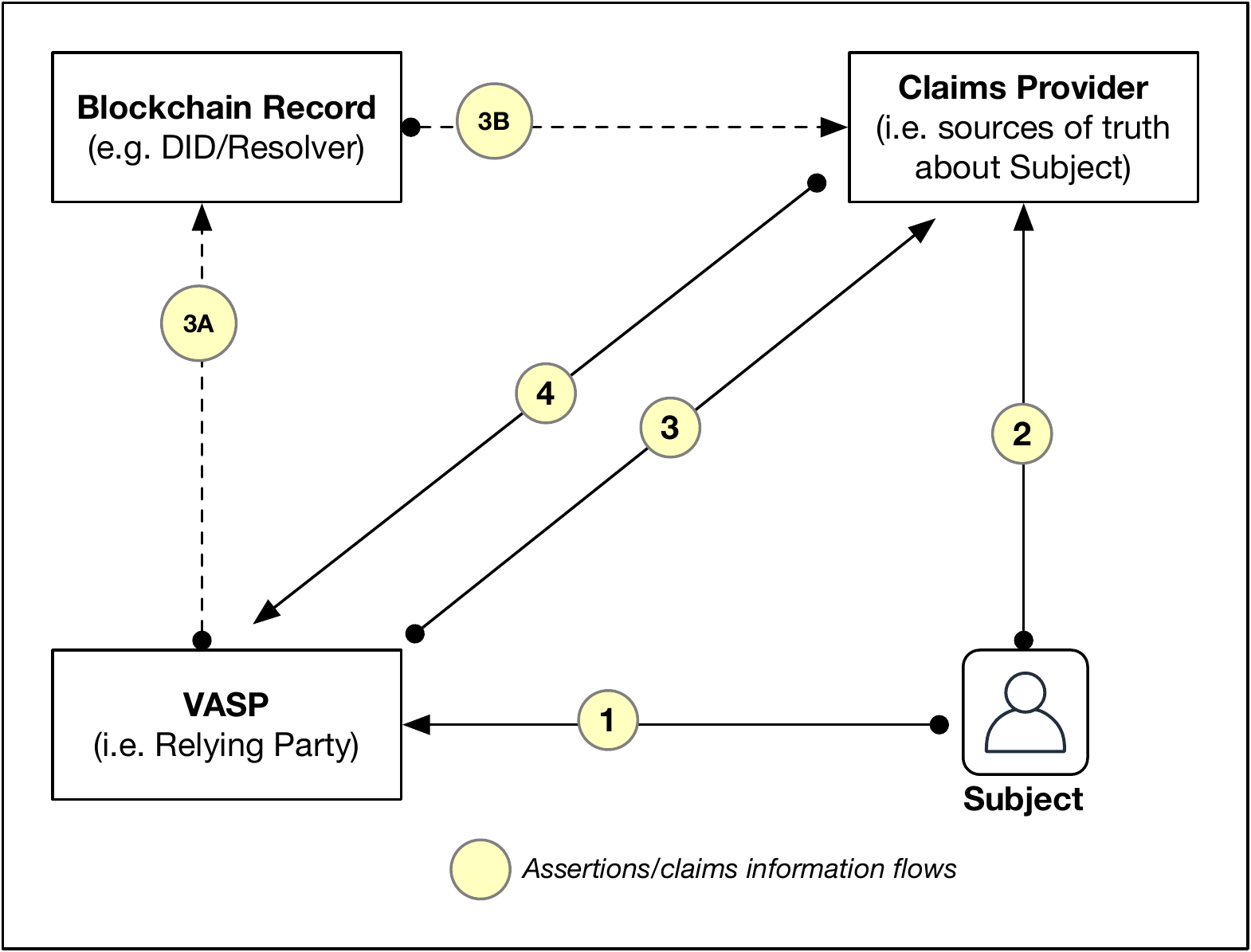}

	%
	%
\caption{The Claims Provider flow}
\label{fig:claimsbasic}
\end{figure}

For VASPs as the relying-party,
the open algorithms approach provides a way for VASPs to be alleviated from
the need to hold the user's data and prove compliance to the data privacy regulations.
Figure~\ref{fig:claimsbasic} summarizes
the basic flow of information (signed claims) from a Claims Provider to the VASP.
In Step~(1) the subject (user) seeks the services of the VASP (e.g. transfer virtual asset).
The subject must indicate to the VASP the identity of the Claims Provider(s)
that can furnish the VASP with information about the subject.
In Step~(2) the subject indicates to the Claims Provider
that the subject consents for the relevant vetted algorithms to be executed by the Claims Provider
in order to yield information or insights sought after by the VASP.
The resulting claims can be static attributes (e.g. ``the user legally lives in city X in country Z'')
or more dynamic insights based on a broader range/type of data over a duration of time
(e.g. ``the user's credit card transactions range for X dollars to Y dollars over the past six month'').
In Step~(3) of Figure~\ref{fig:claimsbasic}
the VASP requests the signed claims from Claims Provider,
which are delivered to the VASP in Step~(4).
By signing the claims,
the Claims Provider implicitly attests to the truthfulness of the statements inside the claims structure.

On the surface it may seem that static attributes may be of primary value for AML/KYC purposes.
However, we believe that dynamic insights based on a broad range of data
may be more useful in the long-term for an ongoing Customer Due Diligence (CDD) program
as discussed in Section~\ref{sec:travelrule}.

\subsection{Linking Claims to Decentralized Identifiers on Blockchains  } 

Although not directly relevant to the problem of deriving claims from data in a privacy-preserving manner,
more recently there have been efforts to use blockchain technology
to enable to the user to better control access to end-points
on the Internet where signed claims may reside.

Referred to as {\em Decentralized Identifiers} (DID)~\cite{W3C-DID-2018},
the basic idea is that the user would ``register'' to the blockchain
a DID-record containing specific end-point configuration information
(e.g. URLs and APIs) 
on the Internet where a requesting party
can obtain information about the user (e.g. location of a store of signed claims).
The record on the blockchain is digitally signed by the user,
indicating that it is the user who self-declares the information about the service endpoint to be true.
Since the user holds the matching private key, 
later if the user seeks to update the DID-record
the user can simply replace it with a newer record (with a newer timestamp).

The idea of a DID as a persistent identifier follows from 
a long history of efforts on persistent and resolvable
digital identifiers on the Internet.
The most prominent of these identifier schemes is 
the {\em Digital Object Identifier} (DOI)~\cite{DOI-ISO-Standard},
with its accompanying {\em Handle} resolver system~\cite{RFC3650}.
Similar in protocol-behavior to the DNS infrastructure,
the DOI and Handle provide for an efficient look-up
of copies of files (e.g. library catalog entries) 
stored at repositories all over the Internet.
Today the DOI and Handle system
have been successfully deployed 
at a wide scale for over a decade 
(e.g. for publications and library records).

Following from Figure~\ref{fig:idpbasic},
an alternate flow using the DID/blockchain approach is shown in Figure~\ref{fig:claimsbasic}
via Steps~(3A) and~(3B).
Here in Step~(1), in addition to the request to the VASP,
the subject provides the VASP with a DID structure (either a public DID or pair-wise DID).
The VASPs resolves the DID value (via the blockchain or DID resolver) in Step~(3A),
which brings the VASP to the correct Claims Provider -- who holds the subject's claims -- in Step~(3B).
As before the Claims Provider responds by delivering the signed claims in Step~(4).

Although the DID/blockchain approach is useful for certain use cases
(e.g. users self-managing their public keys),
in the context of providing relying parties (i.e. VASPs)
with truthful and accurate information about a subject in a privacy-preserving manner,
the role of DIDs remain unclear~\cite{KantaraBSC2017}.

\subsection{Recent VASP  Standardization Efforts} 


Since the issuance of FATF Recommendation~15~\cite{FATF-Recommendation15-2018} there have been
efforts to develop standards to support VASPs
in complying to the FATF and Travel Rule in the context of virtual assets transfers.

The OpenVASP~\cite{OpenVASP-2019} effort borrows from existing modern payments standards,
recasted for the context of cross-VASP exchanges of information.
The goal of OpenVASP is to establish a shared communications protocol for VASPs
to exchange virtual asset transfer information,
as required by the FATP Recommendations.
A related approach is the Travel Rule Information Sharing Architecture (TRISA)~\cite{TRISA-2019}
which seeks to develop a peer-to-peer mechanism for complying with these regulations.
Finally, several organizations in the nascent virtual assets industry
are collaborating to create an  
InterVASP messaging standards~\cite{CallonButler2020}
based on the SWIFT messaging standards in the banking industry.
A detailed analysis of these new proposals is out of scope for the current work.

\section{Privacy-Preserving Claims: Open Algorithms}
\label{sec:opal}

\begin{figure*}[!t]
\centering
\includegraphics[width=1.0\textwidth, trim={0.0cm 0.0cm 0.0cm 0.0cm}, clip]{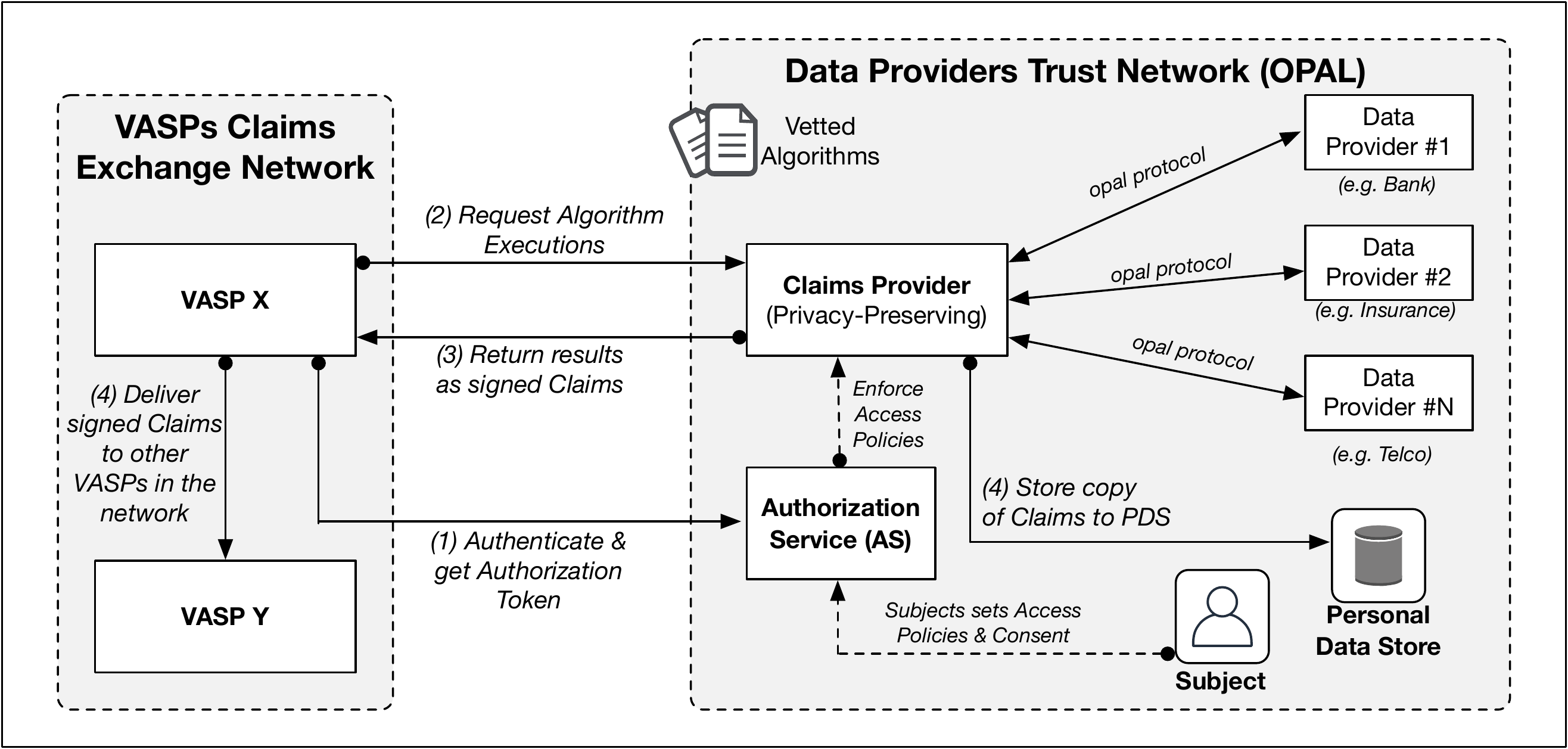}

	%
	%
\caption{The Data Provider Trust Network based on Open Algorithms (after~\cite{HardjonoPentland2018b-SHORT})}
\label{fig:datatrustnetwork}
\end{figure*}

As mentioned previously,
the customer due diligence (CDD) requirements defined by FinCEN
include conducting ongoing monitoring to maintain and update customer
information and to identify and report suspicious transactions.

Today, in order to fulfill the need for ongoing monitoring
of a given subject (person or organization),
data analytics can be performed in order to identify certain trends or to pinpoint certain anomalies.
Extensive data analytics can be performed
only of data is readily accessible.
In reality, however,
today data regarding a subject
is typically stored (siloed) within different institutions across different sectors of industry
(e.g. financial data, health data, social platform behavioral data, etc.).
Furthermore,
each of these data repositories may be operating under different
regulatory jurisdiction that make it difficult 
to combine these data in order to derive better insights~\cite{pentland2015}.
Thus, today we live in a kind of paradox:
huge amounts of digital data are increasingly being accumulated,
but usage for the betterment of individuals and communities are increasingly being hampered
by various constraints.

It is with this backdrop the {\em Open Algorithms} (OPAL)~\cite{pentland-saving-big-data-2014} approach
was first developed at MIT in the context of computational social science,
which sought to obtain better understanding of communities
in the modern digital world.

The open algorithms approach is based on a number of fundamental principles~\cite{HardjonoPentland2019d}:
\begin{itemize}
\item	Data should never leave its repository.
\item	The vetted algorithms are instead transmitted
to the data repository to be executed there.
\item	Only aggregate answers are returned, which
do not permit the re-identification of individuals.
\item	Any algorithm execution yielding a response
that goes deeper or finer-grained than aggregate
results must first obtain explicit consent from the
individuals concerned.
\end{itemize}

A key aspect of the OPAL approach is that 
subject consent by default means {\em permission to execute an algorithm},
which is different from the current industry interpretation of ``consent'' 
(typically meaning permission to export or copy data out of the repository).
Algorithms that are designed to identify individuals (e.g. who satisfy certain criteria)
can only be executed only after explicit consent has been obtain from the individual subject 
(per GDPR~\cite{GDPR}).

The OPAL approach has been piloted in Colombia and Senegal in 2017-2018
in the context of preserving privacy related to research using mobility data 
in those countries~\cite{OPAL-project-status-2018}.
A commercial implementation of OPAL for sharing of insights
among financial industry entities is currently underway.
An extensive discussion of OPAL is out of scope
for this paper and has been treated elsewhere (e.g. see~\cite{HardjonoPentland2018b-SHORT,HardjonoPentland2019d}).

\section{OPAL-Based Data Providers Trust Networks}
\label{sec:opal-vasps}

In order for VASPs to develop a customer due diligence (CDD) program 
that satisfies not only FinCEN and FATF requirements,
but also preserves the privacy of citizens 
-- as required by current privacy regulations (e.g. GDPR and CCPA) and possible future regulations~\cite{Kerry2019a} --
the open algorithms paradigm offers a promising starting point
to derive useful insights that can be conveyed in the form of {\em claims} (or assertions).

More importantly,
for many {\em Data Providers} (data holders) the open algorithms approach
provides the most practical solution that does not require data providers
to give up data -- which is core to the business and which carries its own liabilities.
In many circumstances,
the requesting party (i.e. VASPS) simply need attributes and insights about a subject,
and not raw data about the subject.
As such, for many data holders the open algorithms paradigm may offer them with an avenue to obtain
new revenue streams, 
through the creation of algorithms to match the data in their possession
and by deriving insights (conveyed as claims)) that are relevant to the customer.

Greater effect is created when data providers from distinct industry verticals (e.g. banking \& finance, health, telco, etc.)
collaborate to achieve deeper insights about subjects.
These deeper insights are what a customer due diligence (CDD) programs require
ion the context of virtual assets and VASPs.
We refer to a coalition or consortium of cross-industry data providers
as {\em Data Providers Trust Networks} (see Figure~\ref{fig:datatrustnetwork}).
Similar arrangements have been established in other industries
(e.g. Identity Providers' consortium) for specific purposes
(e.g. share costs for mediated authentication services).
For the nascent VASP industry, collaboration with these data providers trust networks
may be crucial is order to obtain access to these insights based on the open algorithms approach.
In this way VASPs can obtain insights and attributes
based on data of high-provenance,
without needing to resort to data brokers and aggregators.

In Figure~\ref{fig:datatrustnetwork} illustrates the notion of data providers trust networks
supplying insights and attributes in a privacy-preserving manner
using the open algorithms approach.
The interface to the VASP (as the requesting party) is the Claims Provider service.
In Figure~\ref{fig:datatrustnetwork},
before the VASP is permitted to engage the Claims Provider service,
the VASP as a relying party must first be authenticated and be authorized by the {\em Authentication Service} (AS).
This is shown in Step~1 of Figure~\ref{fig:datatrustnetwork}.
The VASP is permitted to choose only from a published list of vetted algorithms.
In Step~2 the VASP submits a request to the Claims Provider.
Responses coming back from the data providers are collated by the Claims Provider
and packaged in the form of a claim or assertion using the relevant format (e.g.~\cite{SAMLcore,Sporny2019}).
The claims or assertions are digitally-signed by the Claims Provider,
and then transmitted to the VASP in Step~3.
A copy of all issued claims or assertions are also placed
in the {\em claims store} of the subject
located, for example, within the {\em Personal Data Store} (PDS)~\cite{Hardjono1996a,openPDS2014PLOS} of the subject.
The copies of signed claims in the subject's PDS claims-store
allows the subject to independently
make use of the claims for other purposes -- which is consistent
with the recommendation of the 2014 WEF report on personal data~\cite{WEF2014}.

An extensive discussion on open algorithms and the data providers trust networks
can be found in~\cite{HardjonoPentland2019d}.

\section{The VASP Claims Exchange Networks: Global Exchange of Claims and Key Information}
\label{sec:VASP-Claims-Network}

\begin{figure*}[!t]
\centering
\includegraphics[width=1.0\textwidth, trim={0.0cm 0.0cm 0.0cm 0.0cm}, clip]{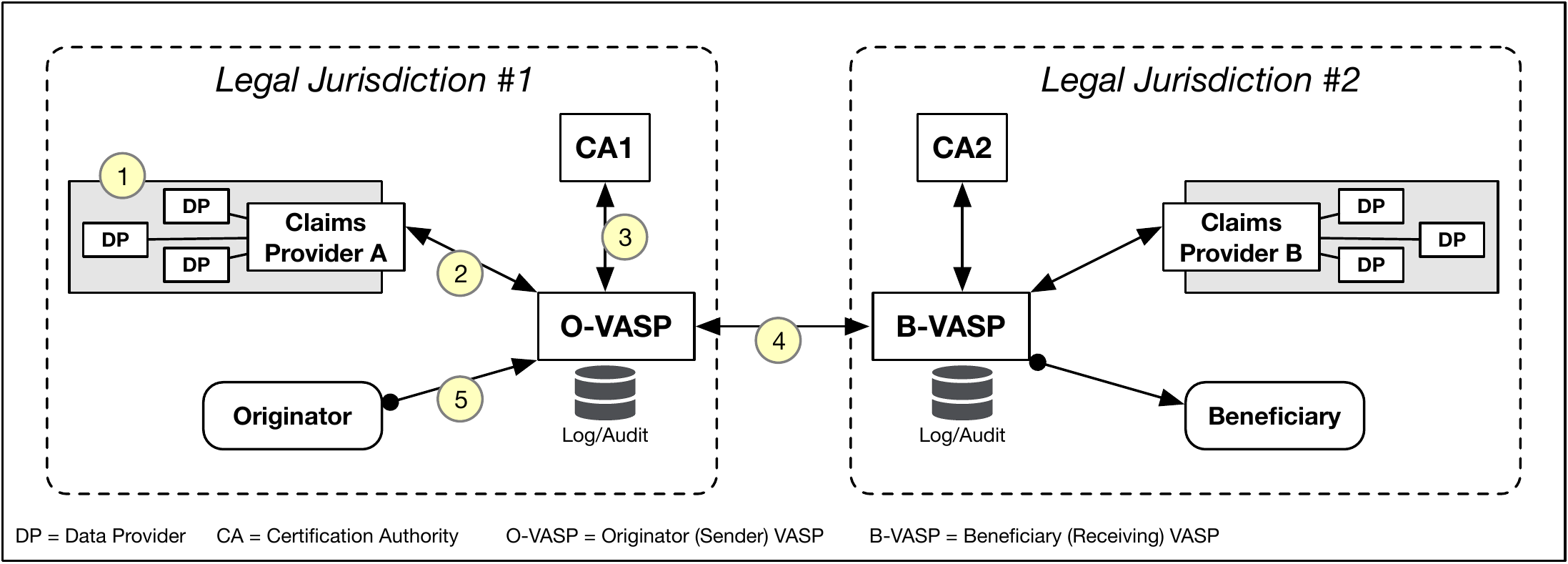}

	%
	%
\caption{VASP Claims Exchange Networks illustrating relationships}
\label{fig:vasptrustnetwork}
\end{figure*}

As mentioned previously,
the Travel Rule requires VASPs to retain information regarding both the
originator and beneficiaries of virtual asset transfers~\cite{FATF-Recommendation15-2018}.
This situation can be challenging when the virtual asset transfer
occurs between VASPs operating under differing legal jurisdictions
(e.g. located in different countries).

To this end we believe that the lessons learned from the evolution of the Internet architecture 
may be beneficial in solving the scaling and interoperability issues related to
VASPs and virtual asset transfers.
More specifically,
communities of VASPs need to form {\em Claims Exchange Networks} (CENs)
for the purpose of (a) sharing of claims generated by local Claims Providers,
and (b) sharing of key ownership information for customers' public keys.
These communities of VASPs that form claims-exchange networks
are akin to autonomous systems in classic IP routing,
where routing domains within an autonomous system
share route reachability information for the benefit of all the member ISPs.

Figure~\ref{fig:vasptrustnetwork} provides an overview of the various relationships
among the entities in a VASP claims-exchange network,
where each relationship is both at the technical level (e.g. APIs and protocols)
as well as at the legal level (e.g. service contracts, practices statements, etc.).
Relationship~(1) of Figure~\ref{fig:vasptrustnetwork}
occurs between the Data Providers (data holders)
with the Claims Provider based on the privacy-preserving 
open algorithms approach~\cite{pentland-saving-big-data-2014,HardjonoPentland2018b-SHORT,HardjonoPentland2019d}.

Relationship~(2) in Figure~\ref{fig:vasptrustnetwork}
is between VASPs and Claims Providers (CP),
where a VASP become a customer of the Claims Provider under a legal service agreement (SLA).
For example,
a signed claim may carry an indication of the provenance or lineage
of the data used by the Data Provider in Relationship~(1) with the Claims Providers.
This gives the VASP some assurance regarding {\em information quality} of the claims,
and thus confidence should the VASP face regulatory scrutiny at a later date about a given subject.
The VASP must retain a copy (log) of all signed claims (regarding all of its originators/beneficiaries)
which it obtained from each Claims Provider with whom it has a business relationship.

Relationship~(3) in Figure~\ref{fig:vasptrustnetwork}
is between VASPs and Certificate Authorities (CA)
as the sources of the key-ownership information regarding private-public keys.
This entails a VASP dealing with multiple CAs
in cases where the originators employ a different CA
from that of the VASP (see~\cite{HardjonoLipton2020a} for a discussion
as to whether a VASP should also be a CA).

Relationship~(4) is between VASPs that may be located within
different legal jurisdictions (e.g. different countries).
One of the core ideas of a Claims Exchange Network for VASPs
is for the establishment of a common set of technical standards (e.g. messaging, key management),
operating procedures, service agreements,
and a shared legal interpretation of the members'
obligations and liabilities.
Include here is a shared agreement regarding the privacy and protection
of data (claims) about originators and beneficiaries~\cite{GDPR}.
Relationship~(5) is between a VASP an its customer (originator or beneficiary).

There are several open challenges with regarding to
the establishment of a claims exchange network of VASPs.
These are discussed below.

\subsection{Assurance of information quality about subjects}

The matter of the {\em provenance of data} becomes crucial
when the origins of claims about a subject need to be
traced back and ascertained.
In the context of the open algorithms paradigm (Figure~\ref{fig:datatrustnetwork})
this means the ability to log, audit and account for the
algorithms and data-sets used by the data providers
(in the data providers trust network)
to derive information about a subject.
The ability for a Claims Provider to account
for each claim or assertion (about a subject) that it issues and signs
reflects the {\em degree of verifiability} of this information.

\subsection{VASPs operations across legal jurisdictions}

The problem of data provenance and the degree of verifiability
of information carried within signed claims
is particularly acute in the case where originators (and Originator-VASPs)
and beneficiaries (and Beneficiary-VASPs) are operating under 
or located within different legal jurisdictions.
The establishment of global industry standards
for information quality assurance for subjects in virtual assets transfers
may have to evolve from national-level standards,
and then later expanded to an international standard through the relevant
standards organizations (e.g. ISO).

When transmitting originator (beneficiary) information or claims -- namely personal data about a subject --
VASPs face a number issues arising due to differing legal jurisdictions.
These include (i) differing or ``mismatched'' data privacy regulations;
(ii) the retention of claims according to local regulations;
(iii) the protection of claims
in transit and in store (data at rest protection).

\subsection{Unmediated decentralized claims exchange networks}

One of the key value propositions of virtual assets (e.g. cryptocurrencies)
is the decentralization of the means to transfer virtual assets~\cite{Bitcoin}.
Here we interpret ``decentralization''
as meaning the ability of an originator (Originator-VASP)
to transfer virtual assets to a beneficiary (Beneficiary-VASP)
without reliance on a centralized party.

Similarly,
networks that exchange claims and key-ownership information
must allow VASPs to exchange claims in an unmediated manner.
A fundamental requirement of claims exchange networks
is the {\em non-repudiation of an exchange of claims}.
That is, an Originator-VASP (Beneficiary-VASP)
must not be able to deny or repudiate that it has transmitted claims
about its customer (subject) to a Beneficiary-VASP (Originator-VASP).

\subsection{Reuse of existing technical standards}

The VASP communities should re-use existing standards in the area of public-key
certificate management, notably the {X.509} standard
that is widely deployed today~\cite{rfc2459,RFC5280-formatted,ISO9594-pubkey} (ISO/IEC 9594-8).
This ensures that VASPs can more easily integrate new services into the existing
security and digital identity infrastructures.
VASPs should also develop their industry-specific Certificate Practices Statement (CPS)
that provides the legal framework for VASPs to determine
risks and liabilities (e.g. due to private-key loss, compromises, etc.).
Various CPS statements exist today from different industries
(e.g.~\cite{SWIFT-CPS-2017,OpenBankingCPS2017,SymantecCPS2013}).

Similarly, VASPs should reuse standards around claims format
(e.g.  {X.509} Attribute certificates~\cite{rfc5755},
the SAML assertions~\cite{SAMLcore}, and the recent Verifiable Claims~\cite{Sporny2019}).

\section{Conclusions}

VASPs face a data problem -- they need truthful information regarding subjects,
such as originators, beneficiaries and other VASPs involved in a virtual asset transfer --
as required by the FATF Recommendations~15 and the Travel Rule.
In this paper we have proposed the Open Algorithms approach
that provides a way for data providers in various industry verticals
(e.g. finance, healthcare, telecom, etc.)
to make insights about subjects
based on their data available in a privacy-preserving manner.
These insights are derived based on the open algorithms principles,
and are delivered via a Claims Provider to the VASP (as the relying party).

In order to scale services to a global level,
we propose that VASP communities form claims-exchange networks
for the purpose of exchanging claims about subjects and key ownership information.
We have discussed a number of challenges today to the establishment
of such networks,
including quality assurance about information contained in claims,
cross-jurisdiction operations of claims-exchange networks,
and the need to reuse or to profile many of the existing technical standards.




\begin{thebibliography}{10}
\providecommand{\url}[1]{#1}
\csname url@samestyle\endcsname
\providecommand{\newblock}{\relax}
\providecommand{\bibinfo}[2]{#2}
\providecommand{\BIBentrySTDinterwordspacing}{\spaceskip=0pt\relax}
\providecommand{\BIBentryALTinterwordstretchfactor}{4}
\providecommand{\BIBentryALTinterwordspacing}{\spaceskip=\fontdimen2\font plus
\BIBentryALTinterwordstretchfactor\fontdimen3\font minus
  \fontdimen4\font\relax}
\providecommand{\BIBforeignlanguage}[2]{{%
\expandafter\ifx\csname l@#1\endcsname\relax
\typeout{** WARNING: IEEEtran.bst: No hyphenation pattern has been}%
\typeout{** loaded for the language `#1'. Using the pattern for}%
\typeout{** the default language instead.}%
\else
\language=\csname l@#1\endcsname
\fi
#2}}
\providecommand{\BIBdecl}{\relax}
\BIBdecl

\bibitem{GDPR}
{European Commission}, ``Regulation {(EU)} 2016/679 of the {E}uropean
  {P}arliament and of the {C}ouncil of 27 {A}pril 2016 on the protection of
  natural persons with regard to the processing of personal data and on the
  free movement of such data ({G}eneral {D}ata {P}rotection {R}egulation),''
  \emph{Official Journal of the European Union}, vol. L119, pp. 1--88, 2016.

\bibitem{CCPA2018}
{California State Legislature}, ``{C}alifornia {C}onsumer {P}rivacy {A}ct
  {(CCPA)} - {AB~375},'' {C}alifornia {C}ivil {C}ode -- Section {1798.100},
  September 2018.

\bibitem{Dreyfuss2018}
\BIBentryALTinterwordspacing
G.~{Chavez-Dreyfuss}, ``Cryptocurrency theft hits nearly \$1 billion in first
  nine months,'' \emph{Reuters Business News}, October 2018. [Online].
  Available:
  \url{https://www.reuters.com/article/us-crypto-currency-crime/cryptocurrency-theft-hits-nearly-1-billion-in-first-nine-months-report-idUSKCN1MK1J2}
\BIBentrySTDinterwordspacing

\bibitem{FATF-Recommendation15-2018}
{FATF}, ``{I}nternational {S}tandards on {C}ombating {M}oney {L}aundering and
  the {F}inancing of {T}errorism and {P}roliferation,'' Financial Action Task
  Force (FATF), {FATF}~{R}evision of {R}ecommendation~{15}, October 2018,
  available at:
  http://www.fatf-gafi.org/publications/fatfrecommendations/documents/fatf-recommendations.html.

\bibitem{FINCEN2014}
{Financial Crimes Enforcement Network (FinCEN) - Department of the Treasury},
  ``{C}ustomer {D}ue {D}iligence {R}equirements for {F}inancial {I}nstitutions
  (31 {CFR} {P}arts 1010, 1020, 1023, 1024, and 1026; {RIN} {1506?AB25}),''
  \emph{Federal Register}, vol.~79, no. 149, August 2014, available at:
  https://www.fincen.gov/sites/default/files/shared/CDD-NPRM-Final.pdf.

\bibitem{FINCEN2019}
\BIBentryALTinterwordspacing
{FinCEN}, ``{A}pplication of {FinCEN's} {R}egulations to {C}ertain {B}usiness
  {M}odels {I}nvolving {C}onvertible {V}irtual {C}urrencies,'' Financial Crimes
  Enforcement Network (FinCEN), {FinCEN}~{G}uidance, May 2019. [Online].
  Available:
  \url{https://www.fincen.gov/sites/default/files/2019-05/FinCEN\%20CVC\%20Guidance\%20FINAL.pdf}
\BIBentrySTDinterwordspacing

\bibitem{WEF2011}
{World Economic Forum}, ``{P}ersonal {D}ata: {T}he {E}mergence of a {N}ew
  {A}sset {C}lass,'' 2011, http://www.weforum.org/reports/
  personal-data-emergence-new-asset-class.

\bibitem{WEF2014}
------, ``{R}ethinking {P}ersonal {D}ata: {A} {N}ew {L}ens for {S}trengthening
  {T}rust,'' May 2014, http://reports.weforum.org/rethinking-personal-data.

\bibitem{Anthem2015}
\BIBentryALTinterwordspacing
R.~Abelson and M.~Goldstein, ``Millions of {A}nthem customers targeted in
  cyberattack,'' \emph{New York Times}, February 2015. [Online]. Available:
  \url{https://www.nytimes.com/2015/02/05/business/hackers-breached-data-of-millions-insurer-says.html}
\BIBentrySTDinterwordspacing

\bibitem{Equifax2017}
\BIBentryALTinterwordspacing
T.~S. Bernard, T.~Hsu, N.~Perlroth, and R.~Lieber, ``Equifax says cyberattack
  may have affected 143 million in the {U.S.}'' \emph{New York Times},
  September 2017. [Online]. Available:
  \url{https://www.nytimes.com/2017/09/07/business/equifax-cyberattack.html}
\BIBentrySTDinterwordspacing

\bibitem{Kerry2019a}
C.~F. Kerry, ``{A} federal privacy law could do better than {C}alifornia?s,''
  Brookings Institution, Report - Center for Technology Innovation, April 2019,
  https://www.brookings.edu/blog/techtank/2019/04/29/a-federal-privacy-law-could-do-better-than-californias/.

\bibitem{FATF-Guidance-2019}
{FATF}, ``{G}uidance for a {R}isk-{B}ased {A}pproach to {V}irtual {A}ssets and
  {V}irtual {A}sset {S}ervice {P}roviders,'' Financial Action Task Force
  (FATF), {FATF}~{G}uidance, June 2019, available at:
  www.fatf-gafi.org/publications/fatfrecommendations/documents/Guidance-RBA-virtual-assets.html.

\bibitem{Hardjono2019b}
T.~Hardjono, ``{C}ompliant {S}olutions for {VASPs},'' May 2019, presentation to
  the FATF Private Sector Consultative Forum (PSCF) 2019, Vienna (6 May 2019).

\bibitem{HardjonoLipton2020a}
T.~Hardjono, A.~Lipton, and A.~Pentland, ``{T}owards a {P}ublic {K}ey
  {M}anagement {F}ramework for {V}irtual {A}ssets and {V}irtual {A}sset
  {S}ervice {P}roviders,'' 2020, {J}ournal of {FinTech} (to appear) --
  Available at {https://arxiv.org/pdf/1909.08607}.

\bibitem{hardjono2019a}
T.~Hardjono, ``{F}ederated {A}uthorization over {A}ccess to {P}ersonal {D}ata
  for {D}ecentralized {I}dentity {M}anagement,'' 2019, {IEEE} {C}ommunications
  {M}agazine (to appear) -- {A}vailable at
  {https://arxiv.org/pdf/1906.03552.pdf}.

\bibitem{SAMLcore}
{OASIS}, ``{A}ssertions and {P}rotocols for the {OASIS} {S}ecurity {A}ssertion
  {M}arkup {L}anguage ({SAML}) {V2.0},'' March 2005, available on
  {http://docs.oasisopen.org/security/ saml/v2.0/ saml-core-2.0-os.pdf}.

\bibitem{Cameron2005identity}
\BIBentryALTinterwordspacing
K.~Cameron, ``The {L}aws of {I}dentity,'' 2005. [Online]. Available:
  \url{https://www.identityblog.com/stories/2005/05/13/TheLawsOfIdentity.pdf}
\BIBentrySTDinterwordspacing

\bibitem{rfc6749}
\BIBentryALTinterwordspacing
D.~Hardt, ``{T}he {OAuth~2.0} {A}uthorization {F}ramework,'' October 2012,
  {RFC6749}. [Online]. Available: \url{http://tools.ietf.org/rfc/rfc6749.txt}
\BIBentrySTDinterwordspacing

\bibitem{UMACORE1.0}
T.~Hardjono, E.~Maler, M.~Machulak, and D.~Catalano, ``{U}ser-{M}anaged
  {A}ccess ({UMA}) {P}rofile of {OAuth2.0} -- {S}pecification {V}ersion
  {1.0},'' Kantara Initiative, Kantara Published Specification, April 2015,
  https://docs.kantarainitiative.org/uma/rec-uma-core.html.

\bibitem{UMACORE2.0}
E.~Maler, M.~Machulak, and J.~Richer, ``{U}ser-{M}anaged {A}ccess ({UMA})
  {2.0},'' Kantara Initiative, Kantara Published Specification, January 2017,
  https://docs.kantarainitiative.org/uma/ed/uma-core-2.0-10.html.

\bibitem{Cook2019}
\BIBentryALTinterwordspacing
T.~Cook, ``You deserve privacy online. here's how you could actually get it,''
  \emph{Time Magazine}, January 2019. [Online]. Available:
  \url{https://time.com/collection-post/5502591/tim-cook-data-privacy/}
\BIBentrySTDinterwordspacing

\bibitem{HardjonoPentland2018b-SHORT}
T.~Hardjono and A.~Pentland, ``{O}pen {A}lgorithms for {I}dentity
  {F}ederation,'' in \emph{{P}roceedings of the {2018} {F}uture of
  {I}nformation and {C}ommunication {C}onference ({FICC}), {Vol.~2}}, K.~Arai,
  S.~Kapoor, and R.~Bhatia, Eds.\hskip 1em plus 0.5em minus 0.4em\relax
  Springer-Verlag, 2018, pp. 24--43.

\bibitem{W3C-DID-2018}
D.~Reed and M.~Sporny, ``{D}ecentralized {I}dentifiers ({DIDs}) {v0.11},'' W3C,
  Draft Community Group Report 09 July 2018, July 2018,
  https://w3c-ccg.github.io/did-spec/.

\bibitem{DOI-ISO-Standard}
{ISO}, ``{D}igital {O}bject {I}dentifier {S}ystem -- {I}nformation and
  {D}ocumentation,'' International Organization for Standardization,
  {ISO~26324:2012}, June 2012, available at:
  http://www.iso.org/iso/catalogue\_detail?csnumber=43506.

\bibitem{RFC3650}
\BIBentryALTinterwordspacing
S.~Sun, L.~Lannom, and B.~Boesch, ``{H}andle {S}ystem {O}verview,'' November
  2003, {RFC3650}. [Online]. Available:
  \url{http://tools.ietf.org/rfc/rfc3650.txt}
\BIBentrySTDinterwordspacing

\bibitem{KantaraBSC2017}
T.~Hardjono and E.~Maler, ``{B}lockchain and {S}mart {C}ontracts {R}eport,''
  Kantara Initiative, Report, June 2017,
  https://kantarainitiative.org/confluence/display/BSC/Home.

\bibitem{OpenVASP-2019}
\BIBentryALTinterwordspacing
D.~Riegelnig, ``{OpenVASP}: An {O}pen {P}rotocol to {I}mplement {FATF's}
  {T}ravel {R}ule for {V}irtual {A}ssets,'' November 2019. [Online]. Available:
  \url{https://www.openvasp.org/wp-content/uploads/2019/11/OpenVasp\_Whitepaper.pdf}
\BIBentrySTDinterwordspacing

\bibitem{TRISA-2019}
\BIBentryALTinterwordspacing
CipherTrace, ``{T}ravel {R}ule {I}nformation {S}haring {A}rchitecture for
  {V}irtual {A}sset {S}ervice {P}roviders {(TRISA)} -- {V}ersion~5,'' December
  2019. [Online]. Available:
  \url{https://ciphertrace.com/wp-content/uploads/2019/08/TRISA-Enabling-FATF-Travel-Rule-V4.pdf}
\BIBentrySTDinterwordspacing

\bibitem{CallonButler2020}
\BIBentryALTinterwordspacing
L.~{Callon-Butler}, ``{Crypto Exchanges Need Common Messaging to Comply With
  Travel Rule},'' \emph{CoinDesk}, February 2020. [Online]. Available:
  \url{https://www.coindesk.com/crypto-exchanges-need-common-messaging-to-comply-with-travel-rule}
\BIBentrySTDinterwordspacing

\bibitem{pentland2015}
A.~Pentland, \emph{{S}ocial {P}hysics: {H}ow {S}ocial {N}etworks {C}an {M}ake
  {U}s {S}marter}.\hskip 1em plus 0.5em minus 0.4em\relax Penguin Books, 2015.

\bibitem{pentland-saving-big-data-2014}
------, ``{S}aving {B}ig {D}ata from {I}tself,'' \emph{{S}cientific
  {A}merican}, pp. 65--68, August 2014.

\bibitem{HardjonoPentland2019d}
T.~Hardjono and A.~Pentland, ``{MIT} {O}pen {A}lgorithms,'' in \emph{{T}rusted
  {D}ata - {A} {N}ew {F}ramework for {I}dentity and {D}ata {S}haring},
  T.~Hardjono, A.~Pentland, and D.~Shrier, Eds.\hskip 1em plus 0.5em minus
  0.4em\relax {MIT} {P}ress, 2019, pp. 83--107.

\bibitem{OPAL-project-status-2018}
\BIBentryALTinterwordspacing
{OPAL~Project}, ``{OPAL}: {S}tatus and {P}lans {2018-19},'' OPAL~Project,
  {S}tatus {R}eport, May 2018. [Online]. Available:
  \url{https://www.opalproject.org/general-overview}
\BIBentrySTDinterwordspacing

\bibitem{Sporny2019}
M.~Sporny, D.~Longley, and D.~Chadwick, ``{V}erifiable {C}redentials {D}ata
  {M}odel {1.0},'' {W3C}, {W3C} {R}ecommendation, November 2019, available at
  {https://www.w3.org/TR/verifiable-claims-data-model}.

\bibitem{Hardjono1996a}
T.~Hardjono and J.~Seberry, ``{S}trongboxes for {E}lectronic {C}ommerce,'' in
  \emph{Proceedings of the Second {USENIX} Workshop on Electronic
  Commerce}.\hskip 1em plus 0.5em minus 0.4em\relax Berkeley, CA, USA: USENIX
  Association, 1996.

\bibitem{openPDS2014PLOS}
Y.~A. {de~Montjoye}, E.~Shmueli, S.~Wang, and A.~Pentland, ``{openPDS}:
  {P}rotecting the {P}rivacy of {M}etadata through {SafeAnswers},'' \emph{PLoS
  ONE 9(7)}, pp. 13--18, July 2014,
  https://doi.org/10.1371/journal.pone.0098790.

\bibitem{Bitcoin}
\BIBentryALTinterwordspacing
S.~Nakamoto, ``{Bitcoin: A Peer-to-Peer Electronic Cash System},'' 2008.
  [Online]. Available: \url{https://bitcoin.org/bitcoin.pdf}
\BIBentrySTDinterwordspacing

\bibitem{rfc2459}
\BIBentryALTinterwordspacing
R.~Housley, W.~Ford, W.~Polk, and D.~Solo, ``Internet {X.509} public key
  infrastructure certificate and crl profile,'' January 1999, {RFC2459}.
  [Online]. Available: \url{http://tools.ietf.org/rfc/rfc2459.txt}
\BIBentrySTDinterwordspacing

\bibitem{RFC5280-formatted}
\BIBentryALTinterwordspacing
D.~Cooper, S.~Santesson, S.~Farrell, S.~Boeyen, R.~Housley, and W.~Polk,
  ``{I}nternet {X.509} {P}ublic {K}ey {I}nfrastructure {C}ertificate and
  {C}ertificate {R}evocation {L}ist ({CRL}) {P}rofile,'' May 2008,
  {IETF}~{S}tandard~{RFC5280}. [Online]. Available:
  \url{http://tools.ietf.org/rfc/rfc5280.txt}
\BIBentrySTDinterwordspacing

\bibitem{ISO9594-pubkey}
{ISO}, ``{I}nformation {T}echnology -- {O}pen {S}ystems {I}nterconnection --
  {T}he {D}irectory -- {P}art~8: {P}ublic-{k}ey and {A}ttribute {C}ertificate
  {F}rameworks,'' International Organization for Standardization,
  {ISO/IEC~9594-8:2017}, February 2017.

\bibitem{SWIFT-CPS-2017}
{SWIFT}, ``{SWIFT} {Q}ualified {C}ertificates for {E}lectronic {S}eals --
  {C}ertification {P}ractice {S}tatement,'' {Symantec Inc.}, Certificate
  Practices Statement, October 2017, https://www.swift.com/pkirepository.

\bibitem{OpenBankingCPS2017}
{Trustis}, ``Open banking certificate policy,'' Open Banking, Certificate
  Policy v1.0. (T-0328-001-GH-001), 2017,
  http://ob.trustis.com/production/policies/.

\bibitem{SymantecCPS2013}
{Symantec}, ``{S}ymantec {S}hared {S}ervice {P}rovider {C}ertification
  {P}ractice {S}tatement,'' {Symantec Inc.}, Certificate Practices Statement
  Version 1.14, April 2013,
  https://www.symantec.com/content/en/us/about/media/repository/ssp-cps.pdf.

\bibitem{rfc5755}
\BIBentryALTinterwordspacing
S.~Farrell, R.~Housley, and S.~Turner, ``An internet attribute certificate
  profile for authorization,'' January 2010, {RFC5755}. [Online]. Available:
  \url{http://tools.ietf.org/rfc/rfc5755.txt}
\BIBentrySTDinterwordspacing

\end{thebibliography}


\end{document}